# Ion-Exchange Doping of Semiconducting Single-Walled Carbon Nanotubes

*Angus Hawkey, Aditya Dash, Xabier Rodríguez-Martínez, Zhiyong Zhao, Anna Champ, Sebastian Lindenthal, Michael Zharnikov, Martijn Kemerink, Jana Zaumseil\**

Angus Hawkey, Xabier Rodríguez-Martínez, Zhiyong Zhao, Sebastian Lindenthal, Michael Zharnikov, Jana Zaumseil
Institute for Physical Chemistry, Heidelberg University, 69120 Heidelberg, Germany

Aditya Dash, Martijn Kemerink
Institute for Molecular Systems Engineering and Advanced Materials, Heidelberg University, Im Neuenheimer Feld 225, 69120 Heidelberg, Germany

Anna Champ
Department of Chemistry, Columbia University, New York, New York 10027, USA

E-mail: zaumseil@uni-heidelberg.de

**Abstract**

Semiconducting single-walled carbon nanotubes (SWCNTs) are a promising thermoelectric material with high power factors after chemical p- or n-doping. Understanding the impact of dopant counterions on charge transport and thermoelectric properties of nanotube networks is essential to further optimize doping methods and to develop better dopants. Here, we utilize ion-exchange doping to systematically vary the size of counterions in thin films of small and large diameter, polymer-sorted semiconducting SWCNTs with $AuCl_3$ as the initial p-dopant and investigate the impact of ion size on conductivity, Seebeck coefficients and power factors. Larger anions are found to correlate with higher electrical conductivities and improved doping stability, while no significant effect on the power factors is found. Importantly, the effect of counterion size on the thermoelectric properties of dense SWCNT networks is not obscured by morphological changes upon doping. The observed trends of carrier mobilities and Seebeck coefficients can be explained by a random resistor model for the nanotube network that accounts for overlapping Coulomb potentials leading to the formation of an impurity band whose depth depends on the carrier density and counterion size. These insights can be applied more broadly to understand the thermoelectric properties of doped percolating disordered systems, including semiconducting polymers.

# 1. Introduction

Organic semiconductors have recently become highly interesting materials for low-temperature (< 100°C) thermoelectric applications due to their typically low thermal conductivity, mechanical flexibility, and low toxicity.[1] While semiconducting polymers have attracted the most attention,[2-3] networks of semiconducting single-walled carbon nanotubes (SWCNTs) are another promising candidate.[4] SWCNTs are a quasi-one-dimensional carbon allotrope which can be understood as a graphene sheet rolled into a seamless tube. The roll-up angle and tube diameter are defined by a pair of chiral indices (n,m) and directly determine the electronic (metallic, semiconducting) and optical (absorption and emission) properties of the nanotube.[5] Semiconducting nanotubes exhibit very high charge carrier mobilities and bandgaps that are inversely proportional to their diameter.[6] Even in random networks, where charge transport is limited by hopping, they still show carrier mobilities (2 - 50 $cm^2 V^{-1} s^{-1}$ )[7-9] that are higher than most semiconducting polymers. Due to their one-dimensional nature, they also have been predicted to reach very high thermoelectric figures of merit.[10-11] Similar to conjugated polymers, semiconducting nanotubes offer solution processability, but with the advantage of achieving comparably higher electrical conductivities ($\sigma$, 1000 - 10,000 S $cm^{-1}$) and Seebeck coefficients ($S$, 60 - 100 µV $K^{-1}$ at the maximum thermoelectric power factor), and thus competitive power factors (PF = $S^2\sigma$, 200 - 920 µW $m^{-1}$ $K^{-2}$).[12-14] While single SWCNTs have high thermal conductivities (> 3000 W $m^{-1}$ $K^{-1}$),[15-16] chemically doped dense films of SWCNTs exhibit only low thermal conductivities (1 – 5 W $m^{-1}$ $K^{-1}$).[17] Overall, thermoelectric figures of merit (z$T = S^2\sigma\kappa^{-1}T$, where $\kappa$ is the thermal conductivity) of up to 0.12 - 0.14 have been reported for doped films of semiconducting SWCNTs.[14, 18]

Key to achieving such thermoelectric performance has been the development of methods to separate mixtures of as-produced metallic and semiconducting nanotubes and to increase the semiconducting purity of the obtained SWCNT dispersions and films.[19] Among them, the selective dispersion of semiconducting or even monochiral (i.e., single (n,m) species) SWCNTs with conjugated polymers in organic solvents has emerged as a highly scalable method to produce large amounts of SWCNTs with >99% semiconducting purity.[20-22] Following this polymer-sorting process, the obtained SWCNT dispersions can be processed into dense films by printing,[23] filtration,[24] ultrasonic spraying,[25] or spin-coating[26], and then chemically doped (p-type or n-type) to achieve the desired conductivities and thermoelectric performance.[12, 14]

While much of the work on SWCNT thermoelectrics has focused on optimizing the SWCNT raw material and film processing,[12-13, 27] less attention has been given to the dopant species and impact of the resulting counterions.[9, 28] However, as a consequence of their one-dimensional nature and the low dielectric constant of the surrounding medium, carbon nanotubes experience only limited screening of Coulomb interactions between their mobile charge carriers and nearby counterions.[29] These Coulomb interactions were argued to lead to the formation of local potential wells, which may trap charge carriers and thus limit charge carrier mobility and conductivity.[28] A better understanding of how counterions impact the charge transport of doped carbon nanotubes could help to enhance the performance of SWCNT thermoelectrics and other SWCNT devices, such as electrochemical transistors.[29-31] As SWCNTs and their random networks might be seen as an extreme form of rigid conjugated polymers,[32] they may also provide further insights into the role of counterions in polymer thermoelectrics.

For semiconducting polymers, ion-exchange doping has been recently used to expand the library of dopant counterions and to investigate their impact on charge transport with a range of different electrolytes and polymers.[33-34] When immersing a conjugated polymer in a dopant solution with a mixture of a redox-active dopant and an excess of electrolyte, the dopant undergoes reduction or oxidation resulting in free charge carriers (holes for *p*-type doping, electrons for *n*-doping) in the polymer and associated dopant counterions nearby. Due to the excess concentration of electrolyte, the dopant counterions are then exchanged with the electrolyte anions in case of *p*-doping, which will be considered here. By exchanging the dopant counterions with anions of increasing size (here defined as the shortest distance to the ionic center of mass of density functional theory (DFT)-optimized ions) the Coulomb interaction between the semiconducting polymer and counterion can be tuned. However, the disruption of the polymer morphology upon introduction of different anions can overshadow the impact of different Coulomb interactions on charge transport.[35] When controlling for film morphology, only a weak dependence of the thermoelectric properties of semiconducting polymers on anion size was found.[36]

The general importance of the counterion size was exemplified through the use of very large perfluorinated dodecaborane (DDB) clusters as chemical *p*-dopants for the semiconducting polymer poly(3-hexylthiophene-2,5-diyl) (P3HT). In this case, the large clusters spatially separated the counterions from the polarons, thus increasing delocalization of charge carriers and hole mobility with increasing counterion size.[37-38] For networks of SWCNTs the power factors were also reported to increase from ~700 µW m$^{-1}$ K$^{-2}$ for doping with the small molecule F$_4$TCNQ (2,3,5,6-tetrafluoro-7,7,8,8-tetracyanoquinodimethane) to 920 µW m$^{-1}$ K$^{-2}$ for doping with the largest DDB dopant (DDB-F$_{72}$).[28] Compared to semiconducting polymers, where the associated morphological changes upon doping often make it difficult or ambiguous to clearly identify the effects of counterion size on charge transport, the length of SWCNTs (0.5 – 1.5 µm), strong van-der Waals forces between nanotubes, and the porosity of a random network should prevent or at least reduce significant structural changes upon ion-insertion during doping. This property makes semiconducting SWCNT networks ideal systems for a systematic study of the impact of anion size on charge transport and thermoelectric properties in percolating disordered systems.

Here, we report ion-exchange *p*-doping of polymer-sorted semiconducting SWCNT networks with small (0.76 nm) and large (1.17-1.55 nm) diameters and hence different bandgaps. We employ AuCl$_3$ as the initial *p*-type dopant and a range of different electrolytes with anions of different sizes to quantitatively evaluate their impact on charge transport and thermoelectric properties of the nanotube networks as well as doping stability. A numerical model based on a random resistor network is applied to understand the relationship between anion size and changes in conductivity and Seebeck coefficient beyond simple charge localization.

## 2. Results
## 2.1. AuCl$_3$ Doping of SWCNT films

To investigate chemical doping of SWCNT thin films, large and small diameter semiconducting nanotubes (**Figure 1**a) were dispersed by shear-force mixing of either CoMoCAT or RN-220 plasma torch nanotube raw material and sorted by selective polymer-wrapping with the polyfluorene copolymer PFO-BPy in toluene as described previously.[22] Excess wrapping polymer was removed by filtration. The selective dispersion of the CoMoCAT SWCNTs with PFO-BPy produces nearly monochiral (6,5) SWCNTs with a diameter of 0.76 nm and a bandgap of 1.27 eV as confirmed by the absorption spectrum (see Figure S1a) with a single narrow E$_{11}$ transition at 997 nm and the E$_{22}$ transition at 576 nm. Dispersing plasma torch (PT) SWCNTs with PFO-BPy produces a mixture of semiconducting chiralities with a diameter of 1.17 – 1.55 nm and a bandgap of 0.70 – 0.88 eV as previously shown.[7, 21] This distribution of semiconducting nanotube species is also evident in the absorption spectra as several overlapping E$_{11}$ and E$_{22}$ absorbance peaks around 1630 nm and 950 nm, respectively (see

Figure S1b). There are no absorption peaks in the region of 600-730 nm, which would correspond to metallic nanotubes.[39] The high semiconducting purity was confirmed by resonant Raman spectra at different excitation wavelengths, which showed only traces of metallic species in both (6,5) and PT SWCNTs (see Figure S2). Note that there is only very little residual PFO-BPy polymer left in the nanotube films as confirmed by the low absorption peaks at 350 nm. The remaining polymer is wrapped around the nanotubes but does not contribute to charge transport due to its large bandgap.[40]

Thin films (20 – 40 nm) of these semiconducting SWCNTs were fabricated by filtering the dispersions on mixed cellulose ester membranes and transferring them to substrates with pre-deposited gold electrodes (see Figure 1b for (6,5) SWCNT films). The resulting films show a high density of nanotubes, while they remain fairly porous (see atomic force micrographs (AFM) in Figure S3). To dope the nanotubes, the films were immersed in acetonitrile (AN) solutions with different concentrations of AuCl$_3$ (10 – 1000 µM). For ion-exchange doping of SWCNTs (see schematic in Figure 1c) the films were immersed in a mixture of AN, AuCl$_3$ and electrolytes with anions of different sizes as shown in Figure 1d.

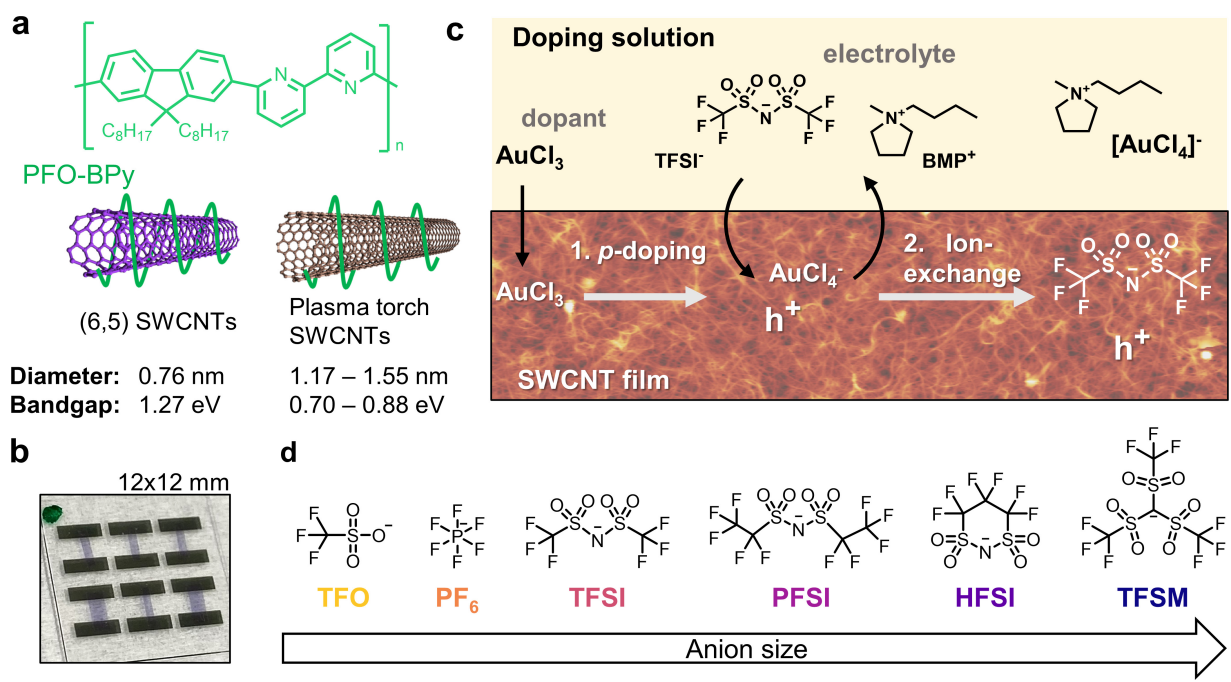

**Figure 1**. (a) Chemical structure of PFO-BPy wrapping polymer and schematic of the polymer-wrapped small-diameter (6,5) SWCNTs (large bandgap) and large-diameter plasma torch nanotubes (small bandgaps). (b) Photograph of filtered and transferred (6,5) SWCNT films

between six pairs of electrodes (spacing 1.5 mm) on glass substrate. (c) Ion-exchange doping reaction scheme for a SWCNT film doped by $AuCl_3$ and anion exchanged with [BMP][TFSI]. (d) Molecular structures of anions of different size used for ion-exchange.

First, we investigate doping of SWCNT films only with $AuCl_3$, which can strongly *p*-dope even small diameter carbon nanotubes with large bandgaps,[41-42] whereas other commonly used dopants such as $F_4TCNQ$ are only mild *p*-dopants that can be applied to large diameter nanotubes.[27] The differences in the one-dimensional density-of-states (DOS) and bandgaps for small and large diameter nanotubes are displayed in **Figure 2**a for (6,5) SWCNTs and (13,5) SWCNTs as a representative for large-diameter PT nanotubes. Furthermore, $AuCl_3$ was selected as a dopant as it was shown to enable more stable doping compared to $FeCl_3$, which is often used for *p*-doping of semiconducting polymers.[34, 43]

Upon immersion of the nanotube film in $AuCl_3$ solution, the SWCNTs are oxidized and $[AuCl_4]^-$ acts as the counterion to the mobile holes.[44] Films of both small and large diameter SWCNTs were submerged in solutions with increasing concentrations of $AuCl_3$ in AN. Consequently, the first SWCNT exciton absorption peak ($E_{11}$) bleaches increasingly as shown in **Figure 2**b-c. The degree of $E_{11}$ bleaching can be used as a reliable quantitative measure for doping of semiconducting nanotubes.[9, 41] In (6,5) SWCNTs this bleaching is accompanied by the emergence of a trion (charged exciton)[45] absorption peak ($T^+$) at approximately 1150 nm, followed by its gradual disappearance at $AuCl_3$ concentrations over 50 mM as previously observed.[46-47] In large-diameter PT SWCNTs no clear trion peaks appear as their expected absorption overlaps with the broad range of $E_{11}$ peaks.[28, 48] In both (6,5) and PT SWCNTs, the second ($E_{22}$) and third ($E_{33}$) exciton subbands, and the absorption of the PFO-BPy wrapping polymer are bleached too, but only at higher doping levels and to a lesser extent. In contrast to polymer films[34, 36, 49] the doping of porous SWCNT networks (see Figure S4) is nearly instantaneous. Due to the high porosity of the network, the majority of nanotubes are immediately exposed to the dopant as the film is immersed in $AuCl_3$ solution, causing a fast bleach of the $E_{11}$ absorption (997 nm). A further slower reduction of absorption probably arises from doping of nanotube bundles or aggregates with a higher polymer content. Consequently, the doping level of these nanotube films cannot be tuned by immersion time as often used for polymers[34, 36] but only by the $AuCl_3$ concentration.

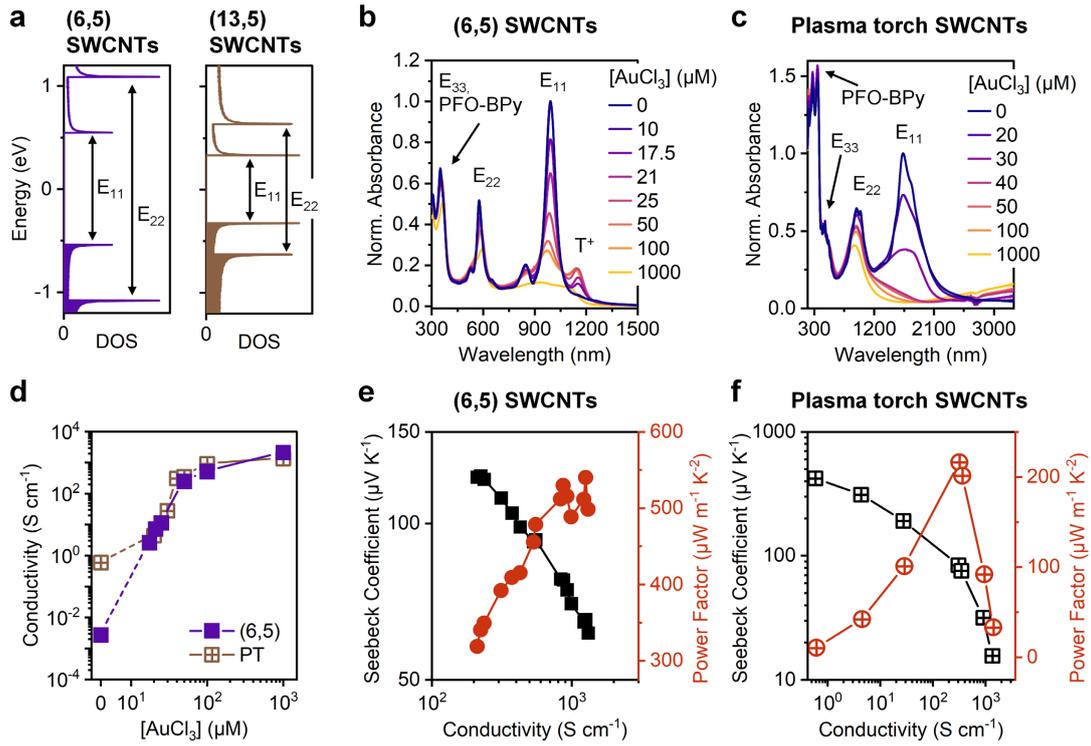

**Figure 2**. (a) Density-of-states (DOS) of (6,5) SWCNTs and (13,5) SWCNTs (as representative of plasma torch SWCNTs). $E_{11}$ and $E_{22}$ represent the main allowed optical transitions. (b,c) Normalized absorption spectra of (6,5) SWCNT films (b) and plasma torch SWCNT films (c) doped by AuCl$_3$/acetonitrile solutions at different concentrations indicating progressive bleaching of $E_{11}$ and $E_{22}$ with increasing doping level. Spectra are normalized to the $E_{11}$ absorbance of the pristine film. T$^+$ represents the trion (charged exciton) absorption. (d) Electrical conductivity of AuCl$_3$-doped (6,5) SWCNT (solid purple squares) and plasma torch SWCNT (crossed brown squares) films depending on dopant concentration. Seebeck coefficients (black squares) and power factors (red circles) for doped films of (6,5) SWCNTs (e) and plasma torch SWCNTs (f) depending on electrical conductivity. Instrumental error bars (typically ca. 10%) are omitted in panels (d, e, f) for clarity.

The electrical conductivities and Seebeck coefficients of the (6,5) and PT SWCNT films were measured before and after doping with increasing concentrations of AuCl$_3$ (see Figure 2d). Before doping (6,5) SWCNTs have a much lower electrical conductivity (2.7×10$^{-3}$ S cm$^{-1}$) compared to PT SWCNTs (0.58 S cm$^{-1}$). Due to their larger bandgap, (6,5) SWCNTs are less sensitive to unintentional *p*-doping in air. They also exhibit intrinsically lower charge carrier mobilities, which increase with the square of the SWCNT diameter.[40, 50] However, after doping with AuCl$_3$, the electrical conductivities are comparable between the two types of films. AFM images (Figure S3) of PT nanotube networks show larger bundles and more aggregates of wrapping polymer compared to networks of (6,5) SWCNTs, which may explain their similar electrical conductivities despite the larger diameter of the PT SWCNTs.[13-14] In both cases the conductivity exceeds 1000 S cm$^{-1}$ when doped with 1 mM AuCl$_3$ and begins to saturate at 100 µM AuCl$_3$ as the E$_{11}$ peaks are completely bleached. The (6,5) SWCNT films in this work show a higher electrical conductivity compared to previously reported thin films of polymer-sorted (7,5) SWCNTs with a similar diameter but produced by ultrasonic spraying.[12] We attribute this difference to the higher density and in-plane alignment of nanotubes in vacuum filtered films. Moreover, shear-force mixing produces longer SWCNTs with a lower number of defects than ultrasonication.[22, 26]

The Seebeck coefficients of the (6,5) SWCNT films are slightly higher (105 - 60 µV K$^{-1}$ for σ = 300 - 1350 S cm$^{-1}$) than those for PT nanotube films (76 – 15 µV K$^{-1}$ for σ = 300 - 1350 S cm$^{-1}$) at a given electrical conductivity but could only be measured in a limited conductivity window. The power factors of doped (6,5) SWCNT films increase up to 540 µW m$^{-1}$ K$^{-2}$ at the maximum of 1 mM AuCl$_3$ used in this study (see Figure 2e). For doped PT SWCNT films the power factor increases to 215 µW m$^{-1}$ K$^{-2}$ and then sharply decreases with further doping (see Figure 2f) due a change of the slope in the Seebeck coefficient versus conductivity dependence at high doping levels, as previously reported for large diameter SWCNTs.[51] Overall the observed behavior of AuCl$_3$-doped SWCNT films is in good agreement with previous reports on the thermoelectric properties of chemically *p*-doped nanotube networks.[12]

## 2.2 Ion-Exchange Doping of SWCNT films

Ion-exchange doping of SWCNT films was first investigated using mixtures of AuCl$_3$ and various concentrations of [BMP][TFSI] electrolyte in AN. When the nanotube film is immersed, the gold chloride *p*-dopes the SWCNTs and the resulting dopant counterion is exchanged with [TFSI]$^-$ (Equation 1).

$$[SWCNT]^+[AuCl_4]^- + [BMP]^+[TFSI]^- \rightleftharpoons [SWCNT]^+[TFSI]^- + [BMP]^+[AuCl_4]^- \quad (1)$$

As derived for ion-exchange in conjugated polymers by Jacobs et al.,[34] the ratio of ionized dopant to electrolyte anion in the film depends on the change in Gibbs free energy for counterion exchange and the ratio of dopant:electrolyte in the doping solution. Hence, the exchange is driven by increasing the concentration of electrolyte anion relative to the dopant in the doping solution. **Figure 3**a shows UV-vis-NIR absorption spectra of (6,5) SWCNT films that were ion-exchange doped by immersion in a mixture of 1 mM AuCl$_3$ and [BMP][TFSI] at concentrations of 0 to 1000 mM in acetonitrile. At each concentration, we observe strong bleaching of the E$_{11}$ and E$_{22}$ peaks. As the [TFSI]$^-$ concentration in the doping solution increases, greater exchange from [AuCl$_4$]$^-$ to [TFSI]$^-$ is confirmed by a reduced absorbance of the [AuCl$_4$]$^-$ metal to ligand charge transfer peak at 228 nm (see Figure 3a, inset). The same trend is observed for films of PT SWCNTs (see Figure S5).

X-ray photoelectron spectroscopy (XPS) of (6,5) SWCNT films doped with AuCl$_3$ confirms the presence of Au and Cl (see Figure S6). No additional XPS peaks are observable when SWCNT films were only treated with [BMP][TFSI], as no doping or ion uptake occurs. However, when SWCNTs films are ion-exchange doped with a 1:100 mM ratio of AuCl$_3$:[TFSI], an F 1s peak emerges, corroborating that [TFSI]$^-$ was incorporated into the nanotube film as the counterion. At the same time, the Au 4f and Cl 2p signals become undetectable, thus indicating near quantitative ion-exchange. These data are in good agreement with the UV-vis-NIR measurements, which also indicate that [AuCl$_4$]$^-$ is almost completely removed from the film at this AuCl$_3$:[TFSI] ratio.

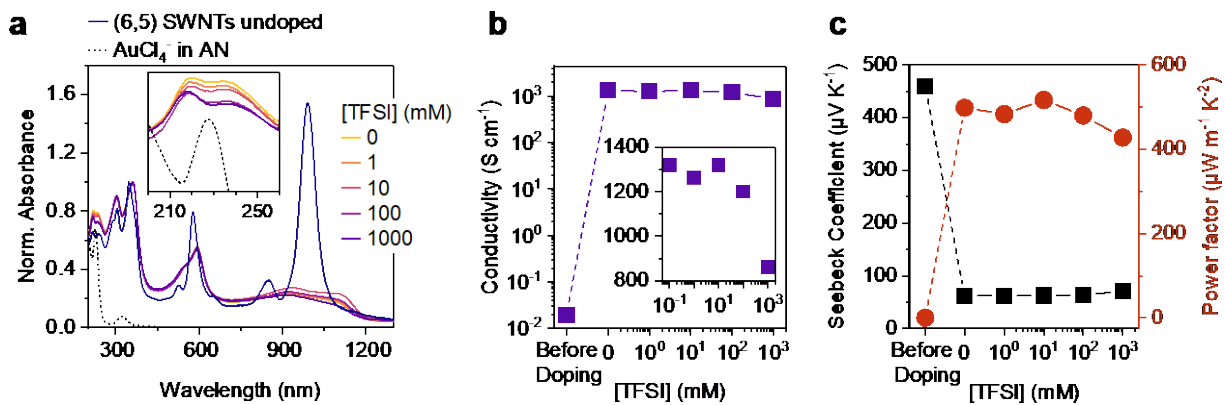

**Figure 3**. (a) UV-vis-NIR absorption spectra of [AuCl4]- in AN (black dashed line), thin films of (6,5) SWCNTs before doping (dark blue line) and after doping with 1 mM AuCl3 and ion-exchanged with [BMP][TFSI] at different concentrations. The inset shows the region where the [AuCl4]- and SWCNT absorbances overlap and progressive removal of [AuCl4]- by ion-exchange is evident. Each spectrum is normalized to its peak at 360 nm. (b) Electrical conductivity (purple squares) with the inset showing zoomed in data in linear-logarithmic scale, (c) Seebeck coefficient (black squares), and power factor (red circles) of (6,5) SWCNT films doped with 1 mM AuCl3 and after ion-exchange with [BMP][TFSI] at different concentrations. Instrumental error bars (typically ca. 6%) are omitted in panels (b, c) for clarity.

After confirming that ion-exchange doping can be applied efficiently to thin films of semiconducting SWCNTs, we investigate their thermoelectric properties (Figure 3b,c). When (6,5) SWCNT films are ion-exchange doped with an increasing concentration of [TFSI]- and fixed concentration of AuCl3, the electrical conductivity remains high (~1300 S cm$^{-1}$) as a result of the charge carrier density being largely unaffected by the ion-exchange process. This notion is corroborated by similar $E_{11}$ absorption values (Figure 3a). The conductivity only decreases to about 860 S cm$^{-1}$ after a AuCl3:[TFSI] ratio of 1:1000 mM has been reached. Similarly, the Seebeck coefficient remains steady (62 µV K$^{-1}$) up to that same AuCl3:[TFSI] ratio (1:1000 mM), in which case the Seebeck coefficient increases slightly (70 µV K$^{-1}$). This behavior may indicate slight de-doping of the (6,5) SWCNT film.

Similar trends in the electrical conductivities and Seebeck coefficients are also observed for the larger diameter PT SWCNTs (see Figure S5), where the electrical conductivity decreases from around 1450 S cm$^{-1}$ to 1070 S cm$^{-1}$ and the Seebeck coefficient increases from 13 µV K$^{-1}$ to 26 µV K$^{-1}$ when increasing the AuCl3:[TFSI] ratio to 1:1000 mM. Notably, the maximum achievable PF for the PT SWCNTs films that were ion-exchange doped with 1:1000 mM

AuCl$_3$:[TFSI] (~75 µW m$^{-1}$ K$^{-2}$) is about seven times lower than that of their (6,5) SWCNT counterparts (~550 µW m$^{-1}$ K$^{-2}$). As shown in Figure 2f, PT SWCNTs are past their peak power factor versus conductivity dependence at an AuCl$_3$ concentration of 1 mM. This implies that small changes in the doping level (i.e., conductivity) may result in abrupt variations of the Seebeck coefficient, which seems to occur at higher concentrations of [TFSI] in the PT SWCNT films (Figure S5c).

Overall, ion-exchange doping in thin films of small and larger diameter SWCNTs is possible when using AuCl$_3$ as the dopant and [BMP][TFSI] as the electrolyte. With greater exchange of [AuCl$_4$]$^-$ to [TFSI]$^-$ the electrical conductivity remains high. As the AuCl$_3$:[TFSI] ratio increases past 1:100 mM, thermoelectric measurements indicate that de-doping might occur (particularly in PT SWCNTs films). Hence, a 1:100 mM AuCl$_3$:anion ratio emerges as the optimum for efficient ion-exchange and high charge carrier densities. Accordingly, this ratio is used as the benchmark ratio for all subsequent experiments.

## 2.3 Influence of Counterion Size on Thermoelectric Properties

Using the established ion-exchange doping of the nanotube networks, we now investigate the dependence of charge transport and thermoelectric properties on counterions size based on the different anions shown in Figure 1d. UV-vis-NIR absorption spectra of (6,5) SWCNT films (see **Figure 4**a) that were ion-exchange doped with different electrolytes show nearly the same bleaching of the E$_{11}$ and also E$_{22}$ absorption, which indicates similar or equal charge carrier densities of approximately 0.3 nm$^{-1}$.[9, 41] Figure 4b displays the measured electrical conductivities versus anion size (defined as the shortest semiaxis of an ellipsoid representing the shape of the molecular ion as described previously).[35-36] At equal absorbance bleach, and thus equal charge carrier density, the conductivity of the SWCNT network increases with counterion size before plateauing. In comparison, the conductivity of semicrystalline polymers following ion-exchange doping does not correlate directly with ion size. Instead, it correlates with the paracrystallinity of the polymer, which is disrupted to different degrees during the doping process depending on the ions used for exchange.[35] However, the conductivity of the nearly amorphous semiconducting polymer IDTBT (poly(indaceno(1,2-b:5,6-b')dithiophene-co-2,1,3-benzothiadiazole)), whose charge transport properties depend less on its morphology, was found to increase with counterion size.[35] As networks of long and rigid SWCNTs are unlikely to undergo morphological changes upon ion-exchange doping, the effect of the counterion size on their charge transport and thermoelectric properties is revealed more clearly as seen in Figure 4b. Interestingly, the Seebeck coefficients of the ion-exchanged (6,5) SWCNT networks decrease with anion size (see Figure S7a) and thus the power factors remain consistently high between 400 and 500 µW m$^{-1}$ K$^{-2}$ for all screened anions (see Figure 4c).

Ion-exchange doped films of PT SWCNTs show a similarly complete absorbance bleach of their E$_{11}$ transition, and partially bleached E$_{22}$ absorption (Figure 4d) indicating comparable or indeed equal p-doping with charge carrier densities of approximately 2 nm$^{-1}$.[9] Note, exchange-doping with [TFO]$^-$ presented an exception as the corresponding E$_{22}$ absorption was bleached less, thus indicating a lower charge carrier density. This difference made direct comparison to PT films doped with the other ions difficult, hence the data was omitted from Figure 4d,e.

As shown in Figure 4e, ion-exchange doped films of PT SWCNTs exhibit a less pronounced rise of the electrical conductivity with anion size compared to the (6,5) SWCNTs. The conductivity increases by only 17% from the smaller anion [PF$_6$]$^-$ to the larger [PFSI]$^-$ anion, compared to a 41% increase for (6,5) SWCNTs. As it is experimentally challenging to exactly match electrical conductivities and power factors of the PT SWCNTs films (which showed a very strong dependence on doping level and conductivity, see Figure 2f), a simple plot of power factor or Seebeck coefficient versus anion size was not possible. Hence, Figure 4f compares the power factors of ion-exchange doped PT SWCNT films with [TFO]$^-$, [TFSI]$^-$, and [PFSI]$^-$ as counterions at various doping levels and electrical conductivities (see Figure S7b for corresponding Seebeck coefficients). In contrast to the results by Murrey et al. for chemically-doped large-diameter SWCNTs,[28] the power factor versus conductivity dependence obtained for all three counterions follows a single and common trend. Hence significant differences in the maximum achievable power factors cannot be assigned to anions of different size. This invariability of the peak power factor versus anion size for both (6,5) and PT SWCNT networks might be attributed to the limited range of anion sizes used here compared to the very large perfluorinated dodecaborane (DDB) clusters used by Murrey et al.[28]

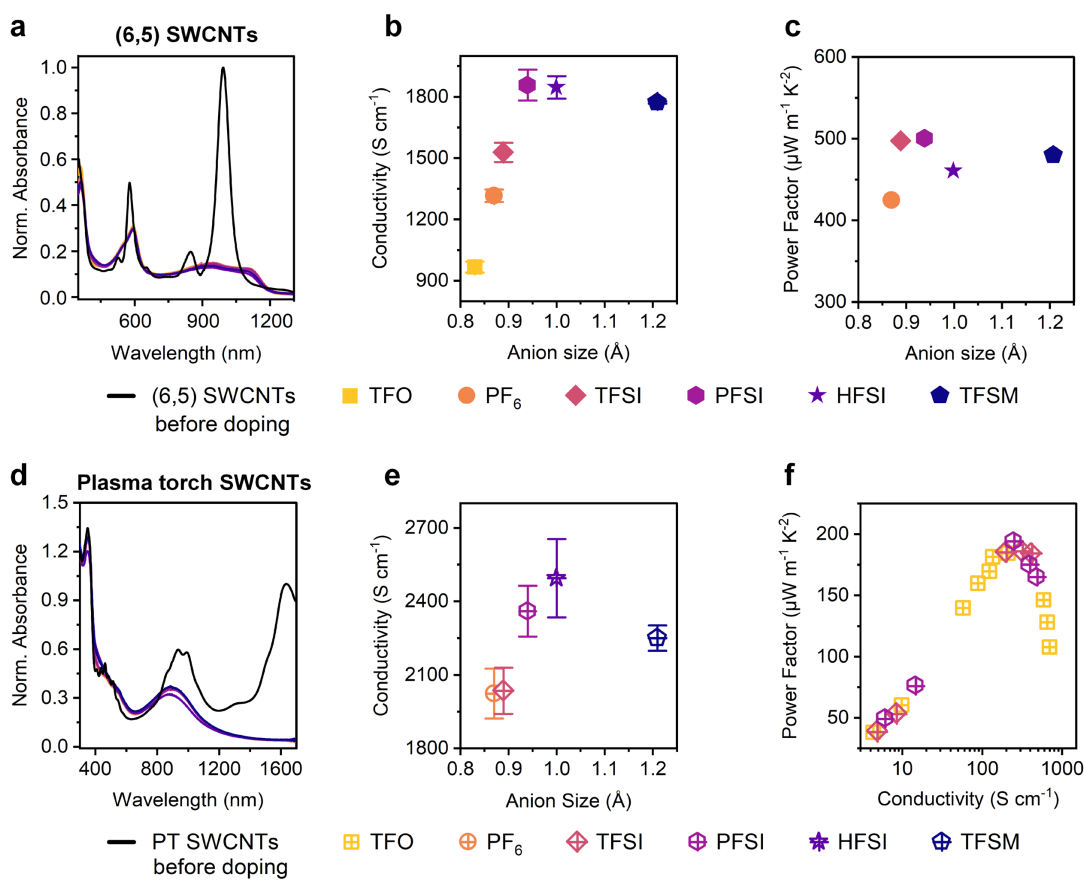

**Figure 4**. UV-vis-NIR absorption spectra of (a) (6,5) SWCNT films and (d) PT SWCNT films doped with 1 mM $AuCl_3$ and ion-exchanged with 100 mM of different anions. Spectra are normalized to the $E_{11}$ absorbance of the pristine film. Corresponding electrical conductivity (b) and power factor (c) of (6,5) SWCNT films versus anion size. (e) Electrical conductivity of ion-exchange doped plasma torch SWCNT films versus anion size. (f) Power factor versus conductivity of PT SWCNT films ion-exchange doped with increasing concentrations of 1:100 $AuCl_3$:anion molar ratio dopant solutions containing [TFO], [TFSI], or [PFSI] as anions. Error bars in panels (b) and (e) result from the standard deviation of 4-6 averaged samples.

To further understand the influence of counterion or anion size on the resulting charge transport, we employ a numerical approach based on a random resistor network model as presented in detail previously,[52] which is supplemented by the solution of the one-dimensional Schrödinger equation using the finite difference method to account for the effect of dopant-induced states. The basic idea behind the model lies in the formation of an impurity band within the bandgap, which is due to the strongly overlapping dopant Coulomb potentials in the high carrier density regime as shown in **Figure 5**a. In general, for low doping levels, the presence of ionized dopant potentials leads to the formation of individual Coulombic traps that give rise to deeply localized states in the density-of-states (DOS) distribution.[53-54] However, in the high doping regime, as in the experiments here, the effective depth of the potential wells is reduced by an energy $\Delta_{overlap}$ due to the strongly overlapping individual Coulomb potentials. This leads to the lowering of the valence band edge from $E_{v,nd}$ to $E_{v,d}$ by $\Delta_{overlap}$, which further facilitates the delocalization of the dopant-induced states, eventually forming an impurity band. The depth of the impurity band ($E_{IB,d}$), *i.e.*, the distance from the band edge ($E_{v,d}$), depends on the dopant density, the dielectric properties and the counterion-nanotube separation distance *d*. This approach and interpretation is in strong contrast to the single trap state approximation used by Eckstein et al.[29, 55] and Murrey et al.[28]

Based on this model Figure 5c and 5d display the simulated results for a constant carrier density. The conductivity rises and Seebeck coefficients decrease with counterion size toward saturation, which quasi-quantitatively fits the experimental data. We attribute the quantitative deviations predominantly to discrepancies in the estimated counterion sizes. In our model, *d* is defined as the distance between the SWCNT, which we approximate as a one-dimensional object, and the dopant ion, which we approximate as a point charge, as depicted in Figure 5b. However, both the finite size of the SWCNT and, more importantly, the actual shape, orientation and position of the counterion affect the Coulombic stabilization and hence the center-to-center distance, especially for non-spherical ions.[35] Unfortunately, there appears to be no consensus on these values. Calculations using different DFT methods lead to strikingly different results.[35, 56] As a first approximation, considering that the ions orient themselves to minimize the distance between their center of charge and the SWCNTs, we use the shortest principal moment of the gyration tensor obtained from DFT-optimized anions, as calculated by Jacobs et al.[35] and Chen et al.[36]

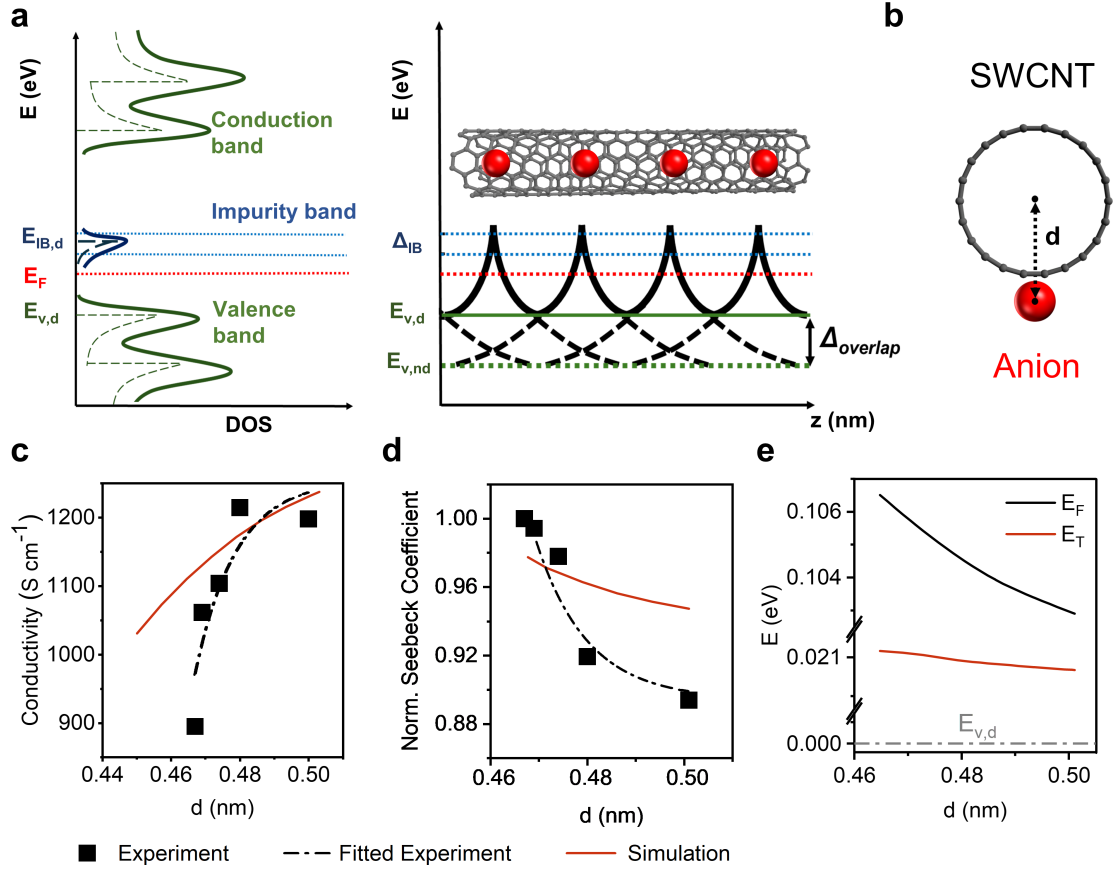

**Figure 5.** (a) Illustration of an impurity band, with a bandwidth $\Delta_{IB}$ and an energetic position $E_{IB,d}$, in the SWCNT DOS due to Coulombic potentials from dopant ions in a highly p-doped SWCNT network. Here, $E_{v,d}$ and $E_{v,nd}$ refer to the valence band edge for the doped and undoped cases, respectively, and the relative difference between these energy levels is denoted as $\Delta_{overlap}$. (b) Schematic view of counterions on SWCNT illustrating nanotube-anion separation $d$. Calculated (c) conductivity and (d) Seebeck coefficients versus $d$, compared with experiments for differently sized counterions for doped (6,5) SWCNT networks at a typical high carrier density of $n = 300$ $e$ μm$^{-1}$. (e) Simulated dependence of transport energy level ($E_T$) and Fermi energy level ($E_F$) on $d$ for $n = 300$ $e$ μm$^{-1}$.

The observed increase in conductivity with ion size is associated with the impurity band moving closer to the valence band edge for larger counter ions. Charge transport in SWCNT networks is largely determined by the junction resistances.[52, 57] In the high doping regime, that is, in the presence of an impurity band, these junction resistances contain a significant contribution from inter-band hops between nanotubes, which become energetically more favored when the impurity bands are shallower. The latter happens when the charge center of the anion is further away from the nanotube (i.e., for large anions). Along the same line of argument, the saturation in conductivity is attributed to the reduced relative change in the impurity band depth with further increasing counterion size. Overall, the simulations indicate that for a fixed doping level or carrier density, larger counterions should improve the SWCNT network conductivity by creating a shallower impurity band.

The Seebeck coefficient is equal to the entropy carried by a single charge or, equivalently, to the heat carried by the moving charge (Peltier coefficient) divided by the temperature $T$. Hence, we calculate the Seebeck coefficient as the difference between transport energy level ($E_T$) and Fermi level ($E_F$) (Equation 2)[58] :

$$S = \frac{k_B}{q} \cdot \frac{(E_T - E_F)}{k_B T} \qquad (2)$$

Here, $k_B$ is the Boltzmann constant and $q$ is the elemental charge. The drop in Seebeck coefficient with increasing counterion size in Figure 5d can accordingly be attributed to the reduced depth of the impurity band. The shallower impurity band can, in turn, be correlated with a decrease in the effective width of the overall DOS. Consequently, for a fixed carrier density, as in our case for the different counter-ion sizes, the narrowing of the DOS pushes the Fermi level towards the center of the DOS, as depicted by the black line in Figure 5e. As is common for percolating disordered systems, the transport energy sits in a region of higher DOS than the Fermi energy and hence is less sensitive to shifts in the lowest energy levels, in this case the impurity band. Hence, the difference ($E_T$ - $E_F$) slowly decreases with increasing counterion size. As such, the small upward shift of $E_T$ and, especially, the larger upward shift of $E_F$ and therefore the drop in Seebeck coefficient corroborates the impurity band moving towards the (higher-lying) valence band. The same argument can be used to explain the increase in conductivity with ion size as the conductivity of the macroscopic percolating network is dominated by thermal activation from the Fermi level to the transport energy, that is, $\sigma \propto e^{-\frac{|E_T - E_F|}{k_B T}}$.

Based on the discussion above and the extended simulated data in Figure S8, a weaker dependence of the transport properties on counterion size can also be predicted for larger diameter nanotubes compared to the (6,5) SWCNTs. While the counterion size range remains the same, its relative contribution to the total distance $d$ from the nanotube center decreases. This prediction is supported by the data in Figure 4e for large-diameter PT SWCNTs, exhibiting a reduced dependence on ion size compared to the small diameter (6,5) SWCNTs for the same range of different anions.

Combining the experimental and theoretical results so far, we find that for similar carrier densities, the electrical conductivity of doped SWCNT networks increases with counterion size while the Seebeck coefficient decreases slightly. In contrast to doping with larger DDB clusters,[28] we do not find an increase in the peak power factor. The applied random resistor model can explain the observed changes of thermoelectric properties with ion size by the formation of an impurity band due to overlapping Coulomb potentials at high doping levels, rather than by localization of the charge carrier due to Coulomb traps as suggested by Murrey et al.[28] The lack of structural changes upon doping of SWCNTs networks make them a useful model system for understanding the effect of counterions on the charge transport and thermoelectric properties of percolating disordered systems. The similarities between SWCNTs and conjugated polymers,[32] specifically their limited screening of Coulomb interactions and fast charge transport along the nanotube or polymer chain, limited by hopping between sites, suggest that the ion size effects on the properties of SWCNTs might be extended to high-mobility polymers in the absence of significant morphological changes upon doping.

## 2.4. Stability of Doped SWCNT Films

Aside from good thermoelectric performance, promising thermoelectric materials should also be stable over time. Unfortunately, the question of stability of thermoelectric materials is often overlooked and few examples of stable *p*-doped or *n*-doped SWCNT thin films have been reported.[59-61] For ion-exchange doped (6,5) SWCNTs stored in inert atmosphere, we find that the conductivity degrades faster with a higher [BMP][TFSI] concentration in the doping solution, i.e., with a greater degree of ion-exchange. The conductivity drops to 15% of its original value after approximately 70 hours when doped with 1:100 mM AuCl$_3$:[TFSI] (see **Figure 6**a). Ion-exchange doped films of large-diameter PT SWCNTs, however, are significantly more stable, retaining 70% of their conductivity after the same time (see Figure 6b).

One cause for the loss in conductivity could be the introduction of defects in the sp² carbon lattice of the nanotubes. Because ion-exchange is an equilibrium process, impurities of AuCl$_3$ may remain in the film after exchange. For semiconducting polymers ion-exchange-doped with FeCl$_3$ and [TFSI], degradation was attributed to residual FeCl$_3$ forming radical species that reacted with the polymer.[62] Residual AuCl$_3$ could reasonably be expected to also form radicals that then react with the nanotube lattice, forming defects which impede charge transport and lower carrier mobilities.[26] Defects in carbon nanotubes can be quantified by the ratio of the defect-related D-mode (~1310 cm$^{-1}$) and the G$^+$-mode (~1589 cm$^{-1}$, related to the sp$^2$ carbon lattice).[63] We find that a significant number of defects are introduced in ion-exchange doped SWCNTs when comparing the Raman D/G$^+$ ratios for (6,5) SWCNTs before doping and a week after doping (see Figure S9). Due to higher lattice strain, small-diameter (6,5) SWCNTs are more reactive compared to large-diameter PT nanotubes[64] and hence are expected to degrade faster as observed in Figure 6a,b. Note that the unintentionally introduced lattice defects and thus changes in the Raman D/G$^+$ ratios of the nanotubes after extended doping are not related to the impurity band discussed above, which is formed by overlapping Coulomb potentials and not structural defects.

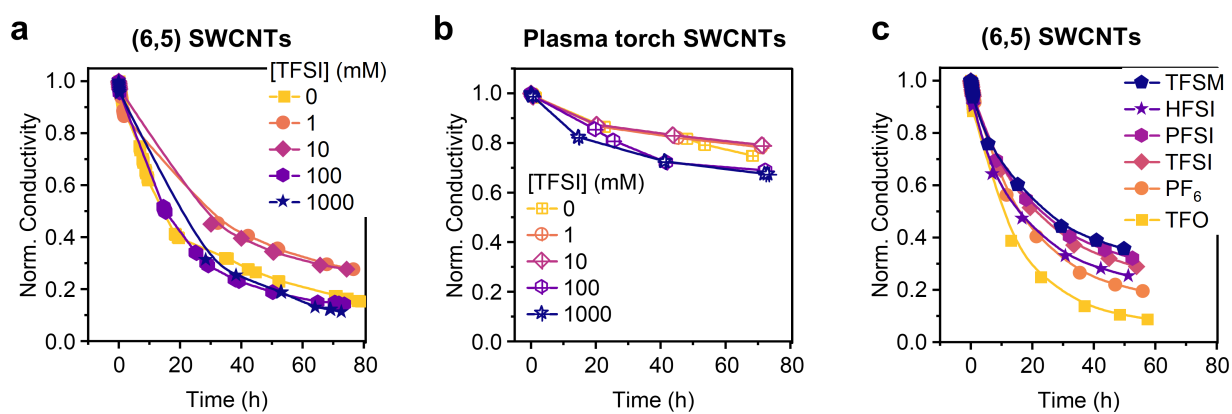

**Figure 6**. Stability of doped SWCNT films stored and measured in a dry nitrogen atmosphere. Conductivity measured over time of (a) (6,5) and (b) plasma torch SWCNT films ion-exchange doped with 1 mM AuCl$_3$ and different concentrations of [BMP][TFSI]. (c) (6,5) SWCNT films ion-exchanged with different anions at 1:100 AuCl$_3$:electrolyte molar ratio.

Figure 6c further shows that the stability of ion-exchanged (6,5) SWCNT networks improves slightly but systematically as the size of the counterion increases. Previously it was shown that residual water, even under glovebox conditions, was sufficient to de-dope *p*-doped organic semiconductors via the water oxidation reaction and formation of hydroxyl radicals. The extent of this reaction depends on the anion used, its hydration, and the ion-cluster size.[65] Large anion sizes should make the water oxidation reaction less favorable. Additionally, the greater diffusion coefficients of smaller anions may contribute to their faster diffusion out of the film, resulting in a slower decrease of the conductivity with larger anions.[66]

Overall, ion-exchange doped large-diameter SWCNTs are more stable than smaller diameter SWCNTs due to the higher reactivity of more strained SWCNTs. The doping stability also correlates with increasing anion size, which may enable further improvement of the long-term stability of chemically doped SWCNT networks by careful selection of dopant and anion combinations.

## 3. Conclusions

We have demonstrated efficient ion-exchange *p*-doping of random networks of polymer-sorted, small and large diameter semiconducting SWCNTs with $AuCl_3$ as the primary dopant and various electrolytes with anions of different size. Ion-exchange doping of SWCNTs enables the introduction of a wide range of counterions without reducing the doping level and hence allows for the investigation of their impact on charge transport and thermoelectric properties. Larger anions result in a significant increase of electrical conductivity at equal doping levels, thus indicating higher charge carrier mobilities. This effect is stronger for small-diameter nanotubes compared to large-diameter nanotubes. However, the peak power factors of the doped networks are not affected by the different counterions. The experimental trends can be rationalized by the overlap of counterion-induced Coulomb potentials leading to the formation of an impurity band at high charge carrier densities that becomes shallower with greater separation between the SWCNT and the charge center of the counterion. The demonstrated impact of the counterion size on the impurity band and hence charge transport in nanotube networks at high doping levels is in contrast to the previously proposed single trap state approximation for doped SWCNTs.[28, 55] Finally, the lack of morphology changes upon ion-exchange doping of SWCNTs networks make them ideal model systems for investigating the effects of different dopant counterions on the charge transport and thermoelectric properties of percolating disordered semiconductors including conjugated polymers. Hence, further studies may provide guidelines for developing dopants and doping methods for improved thermoelectric materials.

## 4. Experimental Section

*Materials:* Salts for ion-exchange doping, 1-Butyl-1-methylpyrrolidinium bis(trifluoromethylsulfonyl)imide ([BMP][TFSI]) (≥ 98.5 %, <0.0400 % water) and tetrabutylammonium hexafluorophosphate ([TBAF][PF$_6$]) (≥ 99.0 %) were purchased from Sigma Aldrich and lithium bis(pentafluoroethanesulfonyl)imide (Li[PFSI]) (>98.0 %), lithium 1,1,2,2,3,3-Hexafluoropropane-1,3-disulfonimide (Li[HFSI]) (>98.0 %), potassium tris(trifluoromethanesulfonyl)methanide (K[TFSM]) (>98.0 %), 1-ethyl-3-methylimidazolium trifluoromethanesulfonate ([EMIM][OTf]) (≥ 98.0 %, ≤ 0.1 % water) from TCI. AuCl$_3$ (≥99.99 %) was purchased from Sigma Aldrich. Doping was performed using anhydrous acetonitrile (<10 ppm water). Salts used for ion-exchange doping were dried under vacuum (0.1 kPa, 60 °C, 48 hours) and transferred to a nitrogen filled glovebox for storage.

*Selective dispersion of s-SWCNTs:* Highly enriched semiconducting (6,5) or plasma torch SWCNTs were prepared by shear force mixing (Silverson L2/Air, 10230 rpm, 20 °C, 72 h) of CoMoCAT SWCNTs (Sigma Aldrich, batch MKCJ7287, 0.4 g L$^{-1}$) or RN-220 SWCNTs (Raymor Industries Inc., batch RNB739-220-161220-A329, 1.5 g L$^{-1}$) in a solution of poly[(9,9-dioctylfluorenyl-2,7-diyl)-alt-(6,6′-(2,2′-bipyridine))] (PFO-BPy, American Dye Source, batch 22B010A1, M$_w$ = 40 kg mol$^{-1}$, 0.5 g L$^{-1}$) in toluene (250 mL). The resulting dispersion was centrifuged twice at 60,000g for 45 min (Beckman Coulter Avanti J26SXP centrifuge) and the semiconducting SWCNT enriched supernatant was collected. Details on the shear force mixing process and overall yield can be found elsewhere.[22, 67] The (6,5) supernatant was additionally filtered through a poly(tetrafluoroethylene) (PTFE) syringe filter (Whatman, pore size 5 μm). Excess polymer was removed by vacuum filtering the dispersion through a PTFE membrane filter (Merck Omnipore, JVWP, pore size 0.1 μm, diameter 25 mm) and soaking in hot toluene (five times for 5 min each, 80 °C). The SWCNTs were redispersed in toluene by bath sonication (30 min) and characterized with UV-vis-NIR absorption spectroscopy. The obtained SWCNTs dispersions were diluted to an optical density of 0.1 cm$^{-1}$ at the E$_{11}$ absorption peak for (6,5) SWCNTs (which according to their molar absorption coefficient[68] corresponds to a concentration of 0.18 mg L$^{-1}$) and 0.025 cm$^{-1}$ at the E$_{22}$ for plasma torch nanotubes. These SWCNT dispersions were finally vacuum filtered (20 mL) through a mixed cellulose esters (MCE) membrane filter (Merck MF-Millipore, VSWP, pore size 0.025 μm, diameter 25 mm) to form a dense film.

*SWCNT film deposition and characterization:* For electrical conductivity and Seebeck measurements, pairs of Cr (2 nm)/Au (20 nm) electrodes with 1.5 mm spacing were evaporated through a shadow mask on clean glass substrates (Schott AF32eco,12×12×0.3 mm for electrical conductivity, 20×25×0.5 mm for Seebeck and electrical conductivity measurements). For optical measurements, clean quartz glass substrates (UQG Optics, FQP-1212) were used. SWCNT films were fabricated by cutting the MCE membranes containing SWCNT films with a razor blade to 0.5-1 mm wide strips. The strips were placed SWCNT-side down between the electrodes, wetted with 2-propanol, hand-pressed onto the substrate to remove air bubbles trapped under the membrane, using aluminum foil to prevent contamination of the films. Finally, they were submerged in acetone (five times for 5 min each) to dissolve the MCE membrane, resulting in a thin film of SWCNTs. No further considerations were made in terms of pressure applied during the transfer process as the SWCNT networks are likely to recover their relaxed state and the resulting thickness was measured afterwards. To account for the sample geometry in the conductivity calculations, film thicknesses were determined with a Bruker DektakXT Stylus profilometer, and film widths and lengths were determined with an Olympus BX51 microscope. All SWCNT films were annealed on a vacuum hotplate for 1 hour at 120 °C and transferred to a dry nitrogen glovebox until used for doping.

*Doping procedure:* Before doping, the SWCNT films were annealed inside a nitrogen-filled glovebox for 15 min at 150 °C. Doping solutions were prepared from stock solutions of $AuCl_3$ (20 mM) and electrolyte (1 M) immediately before doping, typically with 1 mM $AuCl_3$ and 100 mM electrolyte. On a spin-coater, samples were covered by 250 μL doping solution for 5 minutes. Subsequently, they were dried by spinning at 8000 rpm and washed with 250 μL acetonitrile while spinning to remove excess dopant and electrolyte from the surface. Electrical measurements were performed immediately after doping each sample. To maintain consistency within each measurement series (e.g., to study the effect of dopant concentration, or the effect of counterion size), doping was performed on films cut from the same filter membrane and using the same stock solutions of $AuCl_3$ and electrolyte.

*Thermoelectric characterization:* Seebeck coefficients ($S$) were measured in a homebuilt setup by suspending the sample between two Peltier elements and measuring the surface temperature of the glass substrate at each side with PT1000 temperature sensors. Temperature differences across the SWCNT films were calculated under the assumption that the temperature between the Peltier elements changes linearly for small temperature differences. 4-probe current-voltage sweeps (±10 mV) were measured for 6 temperature differences across the SWCNT film within ±1 K using a Keysight B1500A semiconductor parameter analyzer. Thermovoltages were determined from linear fits of the current-voltage sweeps and plot against the temperature difference to give the Seebeck coefficient ($S = \Delta V/\Delta T$). The conductivity (σ) was calculated using the average resistance from the six current-voltage curves and the form factor (lateral dimensions and thickness) of the corresponding sample. The power factor was calculated as PF = $S^2\sigma$.

*Spectroscopic characterization:* Baseline-corrected UV-vis-NIR absorption spectra were collected with a Varian Cary 6000i or Agilent Cary 5000i spectrometer. XPS measurements were performed using a LHS 11 spectrometer (Leybold) and a Mg Kα X-ray source (XR 30, Leybold). SWCNT films were prepared on clean silicon wafer (Siegert wafer GmbH, BW14001) using the method described above. The spectra acquisition was carried out in normal emission geometry with an energy resolution of ~0.9 eV. The X-ray source was operated at a power of 260 W and positioned ~1.5 cm away from the samples. The binding energy scale of the spectra was calibrated to the Au $4f_{7/2}$ peak of clean gold at 84.0 eV. The measurements were performed at room temperature and under UHV conditions, at a base pressure of approximately $8 \times 10^{-10}$ kPa.

*Modelling thermoelectric properties:* A random resistor network model that accounts for both the inter- and intra-tube resistances[52] was expanded with the dopant DOS derived from solving the Schrödinger equation using a finite difference approach to calculate the thermoelectric properties in a doped SWCNT network. In short, we consider a 2D periodic box of 19 μm × 19 μm, where the nanotubes (mean length of 1.4 μm) are randomly distributed (linear density of 15 μm$^{-1}$). The dopant DOS is calculated at a carrier density of $n$ = 300 e μm$^{-1}$ from the distribution of eigenstates of the system's Schrödinger equation, considering the overall dopant potential, $V = \sum_{i}^{N} \frac{-e^2}{4\pi\varepsilon_0\varepsilon_r(d^2+(z-k_i)^2)^{1/2}}$, where $k_i$ represents the $i^{th}$ dopant's position in a system of $N$ dopants, as a perturbation. In the high doping density limit, the modelled dopant DOS is then incorporated as an additional (impurity) band in the calculation of nanotube and junction resistances. The model implements the energetic disorder in the network due to the presence of charged impurities and inherent defects in SWCNTs as a Gaussian distribution of conduction-band-edge energies of the individual nanotubes with a width of 100 meV. All calculations use a relative dielectric constant $\varepsilon_r$ of 3.3, a temperature of 300 K, and an effective hole mass $m_{h,\text{eff}}$ = 0.07·m$_e$.

**Acknowledgements**

This project has received funding from the European Union's Horizon 2020 research and innovation program under the Marie Skłodowska-Curie grant agreement No. 955837 (HORATES). A.C thanks the DAAD-RISE program for support. X.R.M acknowledges funding from the Alexander von Humboldt Foundation. Z. Z. thanks the China Scholarship Council (CSC) for financial support. M.K. thanks the Carl Zeiss Foundation for financial support.

# Supporting Information

# Ion-Exchange Doping of Semiconducting Single-Walled Carbon Nanotubes

*Angus Hawkey, Aditya Dash, Xabier Rodríguez-Martínez, Zhiyong Zhao, Anna Champ, Sebastian Lindenthal, Michael Zharnikov, Martijn Kemerink, Jana Zaumseil\**

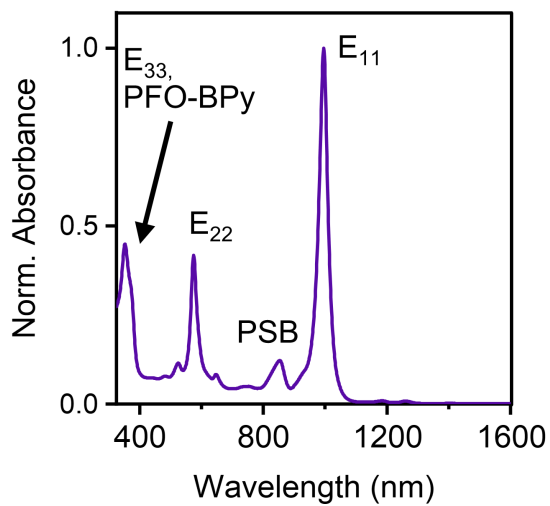 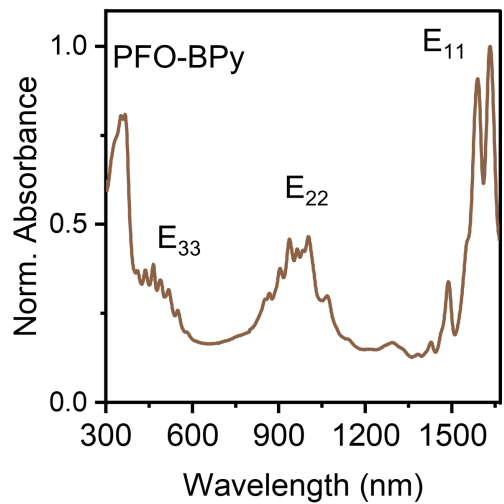

**Figure S1**. UV-vis-NIR absorption spectra of dispersions of PFO-BPy-wrapped (a) (6,5) and (b) plasma torch SWCNTs in toluene normalized to the $E_{11}$ absorbance maximum. PSB – phonon side band.

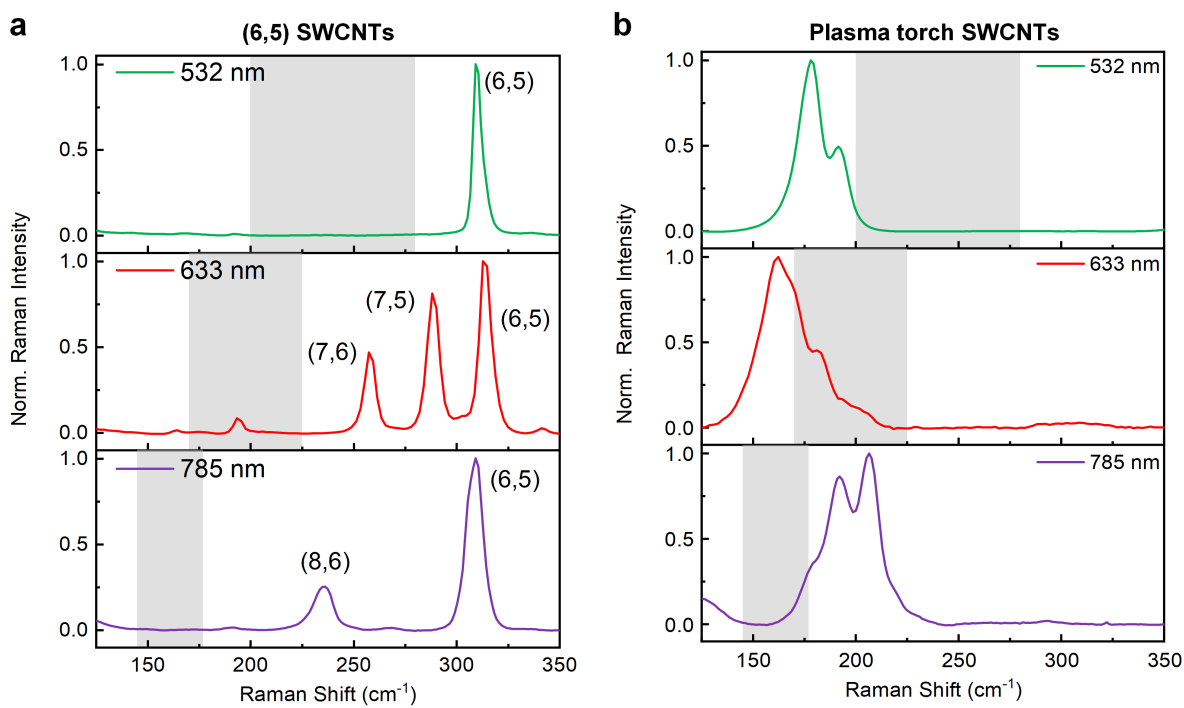

**Figure S2.** Resonant Raman spectra showing the radial breading modes of (a) (6,5) and (b) plasma torch SWCNT films measured at three excitation wavelengths (532 nm, 633 nm, and 785 nm). Note that due to the resonance conditions and high density of the nanotube films, RBM peaks of minority species such as (7,5), (7,6) and (8,6) SWCNTs also become visible, which are barely detectable in the absorption spectra (< 1%, see Figure S1). The gray areas indicate the regions where the RBM peaks of metallic SWCNTs would be expected. Raman spectra were collected with a Renishaw inVia Reflex confocal Raman microscope in the backscattering configuration equipped with a 50x objective (N.A. 0.5). For each spectrum, >1000 individual spectra were collected over an area of 100 μm$^2$ and averaged.

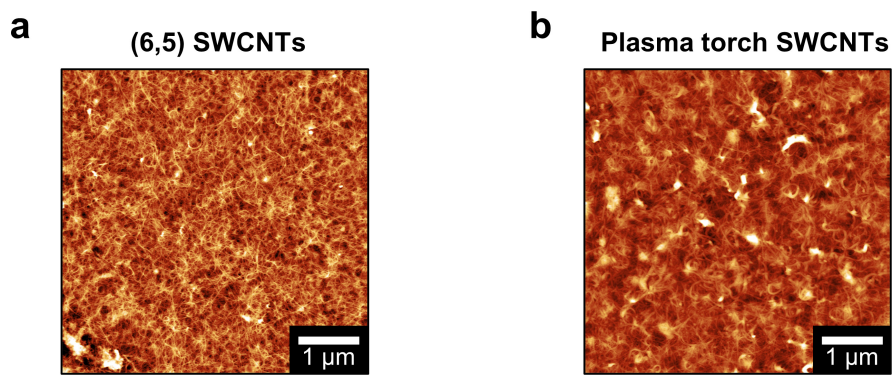

**Figure S3**. Atomic force microscopy (AFM) images of (a) (6,5) and (b) plasma torch SWCNT networks produced by vacuum filtration on mixed cellulose ester membranes and transferred to glass substrates.

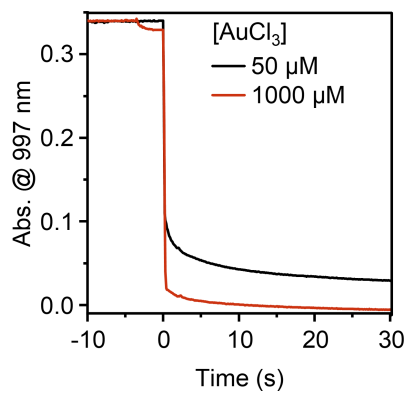

**Figure S4**. Kinetics of doping (6,5) SWCNT films with $AuCl_3$. Absorbance of the $E_{11}$ peak (997 nm) with degree of doping measured over time tracking the bleaching of a SWCNT film on a quartz substrate placed inside a cuvette. At time 0 seconds the cuvette was filled with acetonitrile solutions of $AuCl_3$.

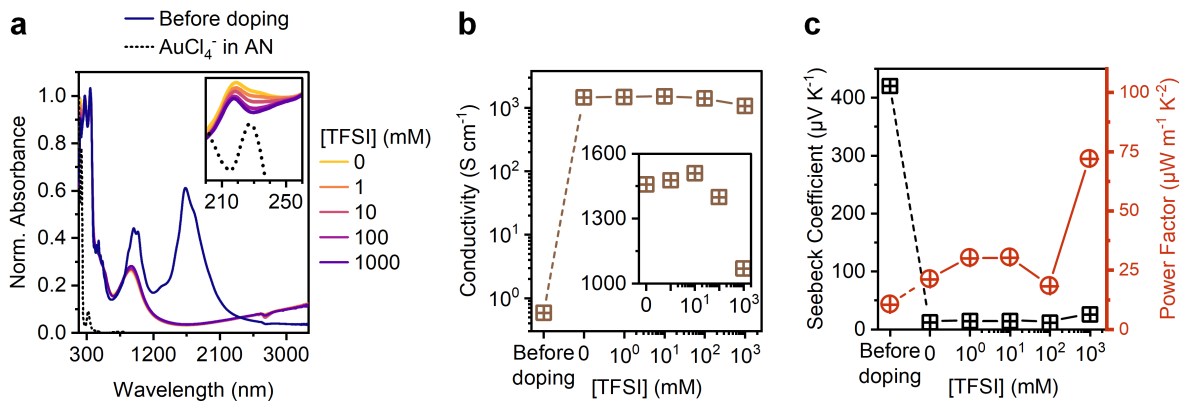

**Figure S5**. (a) UV-vis-NIR absorption spectra of [AuCl$_4$]$^-$ in AN (black dashed line), thin films of plasma torch SWCNT films before doping (dark blue line) and after doping with 1 mM AuCl$_3$ and [BMP][TFSI] at different concentrations. The inset shows the region where the [AuCl$_4$]$^-$ and SWCNT absorbance overlaps and progressive removal of [AuCl$_4$]$^-$ by ion-exchange is evident. Each spectrum is normalized to its peak at 277 nm. (b) Electrical conductivity (brown squares) with the inset showing zoomed in data in linear-logarithmic scale, (c) Seebeck coefficient (black squares), and power factor (red circles) of plasma torch SWCNT films doped with 1 mM AuCl$_3$ and after ion-exchange with [BMP][TFSI] at different concentrations. Instrumental error bars (typically ca. 6 %) are omitted in panels (b, c) for clarity.

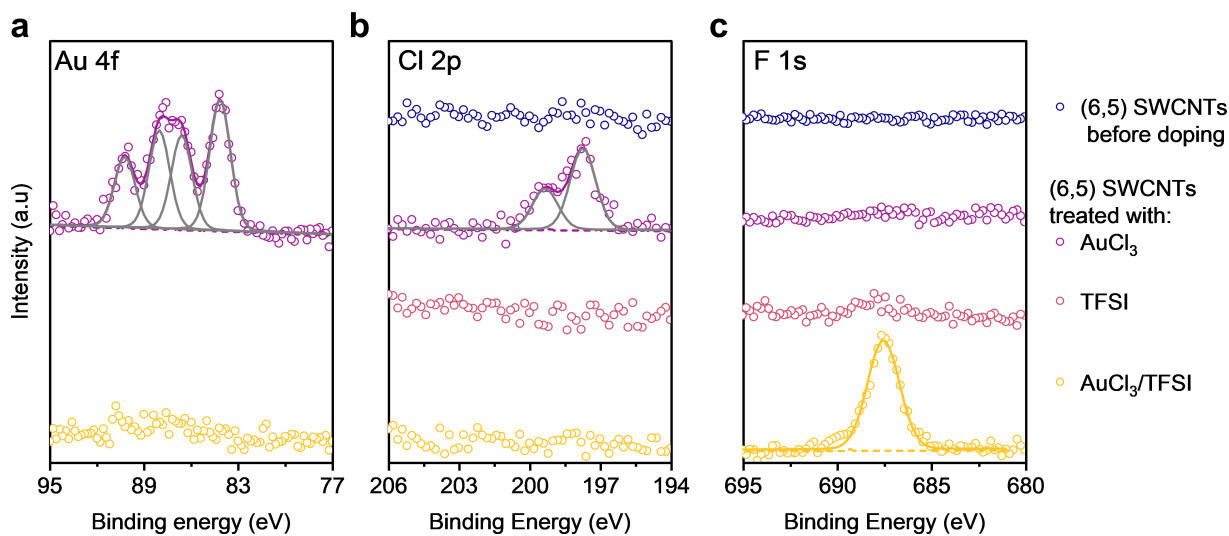

**Figure S6**. XPS photoemission spectra showing (a) Au 4f, (b) Cl 2p and, (c) F 1s of (6,5) SWCNT films, before doping, after treatment with AuCl$_3$ (1 mM), with only [BMP][TFSI] (100 mM), and ion-exchange doped by AuCl$_3$/[BMP][TFSI] (1/100 mM).

S-5

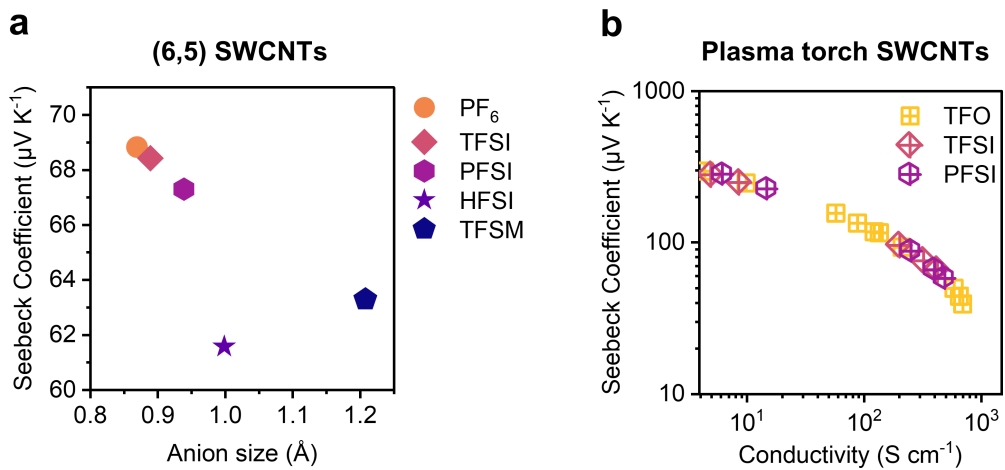

**Figure S7**. (a) Seebeck coefficient versus counterion size for (6,5) SWCNTs. (b) Seebeck coefficient versus conductivity of PT SWCNT films ion-exchange doped with increasing concentrations of 1:100 AuCl$_3$:anion molar ratio dopant solutions containing [TFO], [TFSI], or [PFSI].

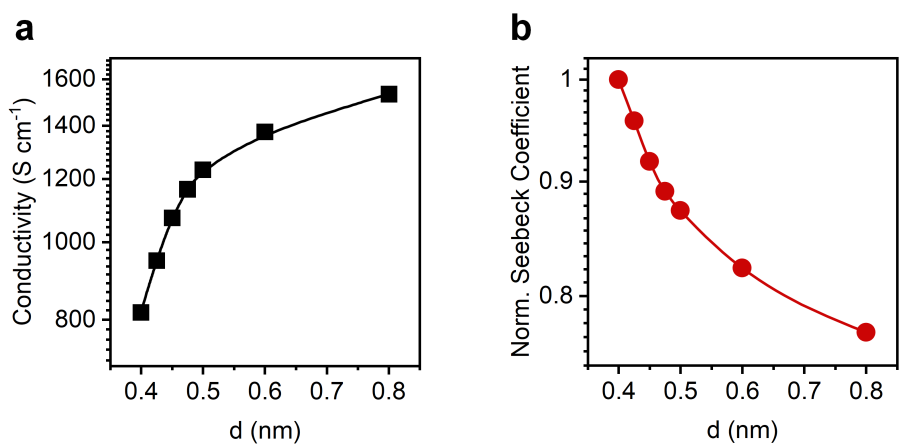

**Figure S8**. Calculated (a) conductivity and (b) Seebeck coefficients versus anion-nanotube separation (*d*) for an extended range of *d* for (6,5) SWCNTs and a charge carrier density of $n = 300$ *e* μm$^{-1}$. The same model parameters were used as in Figure 5.

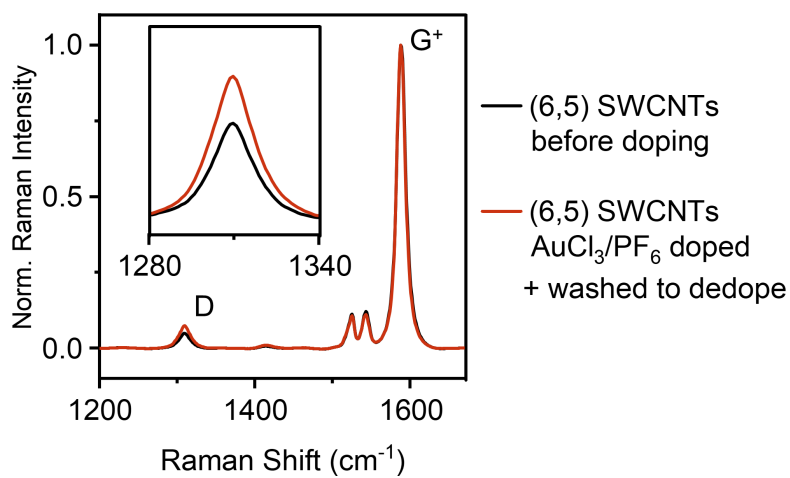

**Figure S9.** Normalized Raman spectra of a (6,5) SWCNT film before doping (black line) and after doping with $AuCl_3/[PF_6]^-$ (1/100 mM) and washing with acetonitrile to de-dope (red line). The inset shows that the intensity of the defect-related D-mode increases after doping and washing to de-dope. Measurements were performed with a 532 nm laser.